%% file: main.tex
\documentclass[final,a4paper,UKenglish,cleveref,autoref,thm-restate]{oasics-v2021}

\title{Towards benchmarking of Solidity verification tools}

\author{Massimo Bartoletti}{University of Cagliari, Italy \and \url{http://blockchain.unica.it} }{bart@unica.it}{https://orcid.org/0000-0003-3796-9774}{Partially supported by project SERICS (PE00000014) and PRIN 2022 DeLiCE (F53D23009130001) under the MUR National Recovery and Resilience Plan funded by the European Union -- NextGenerationEU.}

\author{Fabio Fioravanti}{University of Chieti-Pescara, Italy}{fabio.fioravanti@unich.it}{}{}

\author{Giulia Matricardi}{University of Chieti-Pescara, Italy}{giulia.matricardi@unich.it}{}{}

\author{Roberto Pettinau}{Technical University of Denmark, Denmark}{roberto.pettinau99@gmail.com}{}{}

\author{Franco Sainas}{EPFL, Switzerland}{franco@sainas.me}{}{}

\authorrunning{M. Bartoletti, F. Fioravanti, G. Matricardi, R. Pettinau, and F. Sainas} 



\category{} 

\relatedversion{} 

\nolinenumbers 

\input{macro.tex}

\input{styles.tex}

\makeatletter
\lstdefinestyle{mystyle}{
  basicstyle=%
    \ttfamily
    \lst@ifdisplaystyle\footnotesize\fi
}
\makeatother
\newcommand{\vertasks}{323\xspace}

\begin{document}

\maketitle

\input{abstract.tex}

\input{intro.tex}
\input{related.tex}
\input{benchmark.tex}

\input{solcmc-certora.tex}
\input{conclusions.tex}

\bibliography{main}


\end{document}

%% file: macro.tex
\newcommand{\ifempty}[3]{%
  \ifthenelse{\isempty{#1}}{#2}{#3}%
}

\usepackage{refcount}

\newcommand{\myfootnotetext}[1]{\footnotetext{#1\label{fn:text}%
        \edef\fnmark{\getpagerefnumber{fn:mark}}%
        \edef\fntext{\getpagerefnumber{fn:text}}%
        \ifx\fnmark\fntext\else\ClassWarning{}{footnote mark and text on different pages!}\fi}}

\newcommand{\codefont}{\fontsize{10}{10}\selectfont}
\newcommand{\code}[1]{{\tt\codefont {#1}}}

\newcommand{\solcode}[1]{\lstinline[language=solidity]{#1}}
\newcommand{\cvlcode}[1]{\lstinline[language=cvl]{#1}}

\def\etc{\emph{etc}.\@\xspace}
\newcommand{\Eg}{E.g.\@\xspace}
\newcommand{\eg}{e.g.\@\xspace}
\newcommand{\ie}{i.e.\@\xspace}

\DeclareMathSymbol{:}{\mathord}{operators}{"3A}



%% file: styles.tex
\usepackage[T1]{fontenc}

\usepackage{color}
\usepackage[usenames,dvipsnames]{xcolor}

\usepackage[framemethod=tikz]{mdframed}

\usepackage{multicol}
\usepackage{booktabs}
\usepackage{siunitx}
\usepackage{textcomp}

\usepackage{caption}
\DeclareCaptionType{listing}[Listing][List of Listings] 
\crefname{listing}{Listing}{Listings}  

\usepackage[most]{tcolorbox}

\tcbset{myoneside/.style={
colback=LightGrey,
colframe=gray,
top=0pt,
left=2pt,
right=2pt,
bottom=-5pt,
boxsep=1pt,
halign=flush center,
nobeforeafter,
before skip=-5pt,
after skip=0pt,
boxrule=1pt
}}

\tcbset{mytwosides/.style={
colback=LightGrey,
colframe=gray,
top=0pt,
left=2pt,
right=2pt,
bottom=-5pt,
boxsep=1pt,
sidebyside,
sidebyside align=top,
sidebyside gap=8pt,
halign=flush center,
nobeforeafter,
before skip=-5pt,
after skip=0pt,
boxrule=1pt
}}

\usepackage{pst-func}
\usepackage{amssymb}
\usepackage{graphicx}
\usepackage{svg}

\usepackage{paralist}
\usepackage{environ}

\usepackage{tikz}
\usepackage{forest}
\usetikzlibrary{shadows}
\usetikzlibrary{matrix,chains,positioning,decorations.pathreplacing}
\usetikzlibrary{patterns,calc,fit,arrows}
\usetikzlibrary{shapes.geometric}
\usetikzlibrary{positioning,arrows.meta,backgrounds}

\usepackage{mathtools}
\usepackage{array,multirow}
\usepackage{algorithm}

\usepackage{mdframed}
\usepackage{arydshln}
\usepackage{etoolbox}
\usepackage{xspace} 
\usepackage{dirtytalk} 
 
\usepackage{makecell}
\usepackage{stackengine}
\setstackEOL{\cr} 

\usepackage[inline,shortlabels]{enumitem} 
\newlist{inlinelist}{enumerate*}{1}
\setlist*[inlinelist,1]{%
  label=(\roman*),
}

\usepackage{nicefrac}

\usepackage[normalem]{ulem} 

\usepackage{xifthen}  
\usepackage{ifthen}

\usepackage{cleveref}
\usepackage{bigints}
\usepackage{showkeys}

\usepackage[final,nomargin,inline,index]{fixme} 
\fxusetheme{color}

\FXRegisterAuthor{bart}{anbart}{\color{magenta} {\underline{bart}}}
\FXRegisterAuthor{rob}{anrob}{\color{teal} {\underline{rob}}}
\FXRegisterAuthor{giulia}{angiulia}{\color{blue} {\underline{giulia}}}
\FXRegisterAuthor{ff}{anff}{\color{red} {\underline{ff}}}
\FXRegisterAuthor{franco}{anfranco}{\color{green} {\underline{franco}}}

\hypersetup{
  breaklinks   = true,
  colorlinks   = true, 
  urlcolor     = blue, 
  linkcolor    = blue, 
  citecolor    = red   
}
\hypersetup{final} 

\usepackage{xurl} 

\usepackage{listings}
\definecolor{listingBG}{HTML}{FFFFCB}%
\definecolor{listingFrame}{HTML}{BBBB98}%
\definecolor{listingLineno}{rgb}{0.5,0.5,1.0}%

\definecolor{LightGrey}{rgb}{0.975,0.975,0.975}

\usepackage{listings}
\definecolor{listingBG}{HTML}{FFFFCB}%
\definecolor{listingFrame}{HTML}{BBBB98}%
\definecolor{listingLineno}{rgb}{0.5,0.5,1.0}%

\definecolor{LightGrey}{rgb}{0.975,0.975,0.975}

\lstdefinelanguage{solidity}{
	commentstyle=\color{Gray},
	morecomment=[l]{//},
	morecomment=[s]{/*}{*/},
	classoffset=0,
        escapechar=\$,
	morekeywords={anonymous, assembly, balance, break, call, callcode, case, catch, class, constant, continue, constructor, contract, debugger, default, delegatecall, delete, do, else, emit, event, experimental, export, external, false, finally, for, function, gas, if, implements, import, in, indexed, instanceof, interface, internal, is, length, library, log0, log1, log2, log3, log4, memory, modifier, new, payable, pragma, private, protected, public, pure, push, return, returns, revert, rule, selfdestruct, send, solidity, storage, struct, suicide, super, switch, then, this, throw, transfer, true, try, typeof, using, value, view, while, with, addmod, ecrecover, keccak256, mulmod, ripemd160, sha256, sha3},
	keywordstyle=\color{NavyBlue}\bfseries,
	classoffset=1,
	morekeywords={address, bool, byte, bytes, bytes1, bytes2, bytes3, bytes4, bytes5, bytes6, bytes7, bytes8, bytes9, bytes10, bytes11, bytes12, bytes13, bytes14, bytes15, bytes16, bytes17, bytes18, bytes19, bytes20, bytes21, bytes22, bytes23, bytes24, bytes25, bytes26, bytes27, bytes28, bytes29, bytes30, bytes31, bytes32, enum, int, int8, int16, int24, int32, int40, int48, int56, int64, int72, int80, int88, int96, int104, int112, int120, int128, int136, int144, int152, int160, int168, int176, int184, int192, int200, int208, int216, int224, int232, int240, int248, int256, mapping, string, uint, uint8, uint16, uint24, uint32, uint40, uint48, uint56, uint64, uint72, uint80, uint88, uint96, uint104, uint112, uint120, uint128, uint136, uint144, uint152, uint160, uint168, uint176, uint184, uint192, uint200, uint208, uint216, uint224, uint232, uint240, uint248, uint256, var, void, ether, finney, szabo, wei, days, hours, minutes, seconds, weeks, years},
	keywordstyle=\color{blue},
	classoffset=2,
	morekeywords={block, blockhash, coinbase, difficulty, gaslimit, number, timestamp, msg, data, gas, sender, sig, value, now, tx, gasprice, origin},
	keywordstyle=\color{Plum}\bfseries,
 	classoffset=3,
 	morekeywords={assert,require},
	keywordstyle=\color{red}\bfseries,
}

\lstdefinelanguage{cvl}{
	commentstyle=\color{Gray},
	morecomment=[l]{//},
	morecomment=[s]{/*}{*/},
	classoffset=0,
        escapechar=\$,
	morekeywords={anonymous, assembly, balance, break, call, callcode, case, catch, class, constant, continue, constructor, contract, debugger, default, delegatecall, delete, do, else, emit, event, experimental, export, external, false, finally, for, function, gas, if, implements, import, in, indexed, instanceof, interface, internal, is, length, library, log0, log1, log2, log3, log4, memory, modifier, new, payable, pragma, private, protected, public, pure, push, return, returns, revert,  selfdestruct, send, solidity, storage, struct, suicide, super, switch, then, this, throw, transfer, true, try, typeof, using, value, view, while, with, addmod, ecrecover, keccak256, mulmod, ripemd160, sha256, sha3},
	keywordstyle=\color{NavyBlue}\bfseries,
	classoffset=1,
	morekeywords={address, bool, byte, bytes, bytes1, bytes2, bytes3, bytes4, bytes5, bytes6, bytes7, bytes8, bytes9, bytes10, bytes11, bytes12, bytes13, bytes14, bytes15, bytes16, bytes17, bytes18, bytes19, bytes20, bytes21, bytes22, bytes23, bytes24, bytes25, bytes26, bytes27, bytes28, bytes29, bytes30, bytes31, bytes32, enum, int, int8, int16, int24, int32, int40, int48, int56, int64, int72, int80, int88, int96, int104, int112, int120, int128, int136, int144, int152, int160, int168, int176, int184, int192, int200, int208, int216, int224, int232, int240, int248, int256, mapping, string, uint, uint8, uint16, uint24, uint32, uint40, uint48, uint56, uint64, uint72, uint80, uint88, uint96, uint104, uint112, uint120, uint128, uint136, uint144, uint152, uint160, uint168, uint176, uint184, uint192, uint200, uint208, uint216, uint224, uint232, uint240, uint248, uint256, var, void, ether, finney, szabo, wei, days, hours, minutes, seconds, weeks, years},
	keywordstyle=\color{blue},
	classoffset=2,
	morekeywords={block, blockhash, coinbase, difficulty, gaslimit, number, timestamp, msg, data, gas, sender, sig, value, now, tx, gasprice, origin},
	keywordstyle=\color{Plum}\bfseries,
 	classoffset=3,
 	morekeywords={invariant,rule,assert,satisfy,require},
	keywordstyle=\color{red}\bfseries,
}

%% file: abstract.tex
\begin{abstract}
Formal verification of smart contracts has become a hot topic in academic and industrial research, given the growing value of assets managed by decentralized applications and the consequent incentive for adversaries to tamper with them.
Most of the current research on the verification of contracts revolves around Solidity, the main high-level language supported by Ethereum and other leading blockchains.
Although bug detection tools for Solidity have been 
proliferating almost since the inception of Ethereum, only in the last few years we have seen verification tools capable of proving that a contract respects some desirable properties.  
An open issue is how to evaluate and compare the effectiveness of these tools:
indeed, the existing benchmarks for general-purpose programming languages cannot be adapted to Solidity, given substantial differences in the programming model and in the desirable properties. 
We address this problem by proposing an open benchmark for Solidity verification tools. 
By exploiting our benchmark, we compare two leading tools, SolCMC and Certora, discussing their completeness, soundness and expressiveness limitations.
\end{abstract}

%% file: intro.tex
\section{Introduction}
\label{sec:intro}

The rapid growth of decentralized applications based on blockchain technologies have emphasized the importance of ensuring the security of smart contracts --- the basic building blocks of these applications. The research on smart contracts security has been proliferating since 2016, leading on the one side to the discovery of a variety of attacks, and on the other side to the development of several tools to detect vulnerabilities of smart contracts before they are deployed. Despite the increasing breadth and precision of these analysis tools, attacks to smart contracts have caused financial losses worth several billions of dollars so far, and are unlikely to be eradicated anytime soon. 

A large class of analysis tools for smart contracts are focussed on detecting known vulnerability patterns in contracts code. Even though tools of this type can detect many nefarious bugs, statistically the vast majority of the losses due to real-world attacks are caused by logic errors in the contract code, which cannot be prevented by only checking for fixed vulnerability patterns~\cite{Chaliasos24icse}. 
In this context, a contract can be considered secure when its executions are coherent with some ideal behaviour, even in the presence of adversaries trying to subvert it. 
Only a few tools support this kind of security analysis, allowing developers to specify the ideal properties the contract is expected to satisfy. In this work we focus on SolCMC and Certora, two leading verification tools for contracts written in Solidity, the main smart contract language for Ethereum and EVM-compatible blockchains. Both tools allow the developer to specify desirable contract properties, and use SMT solvers to verify whether the contract satisfies them, showing a counterexample when detecting a violation. 
Although both tools have been independently tested by their developers~\cite{solcmc-cav22-artifact,certora-openzeppelin22}, no public comparison exists so far to assess their effectiveness and limitations in practice.

Our long-term goal is a comprehensive, publicly available benchmark to evaluate the effectiveness of verification tools for Solidity contracts. As an initial step towards this goal, in this paper we present a benchmark comprising \vertasks verification tasks, each one made of a Solidity contract and a property it is expected to satisfy.%
\footnote{\url{https://github.com/fsainas/contracts-verification-benchmark}}
%
A crucial component of our benchmark is a manually crafted ground truth of the verification tasks, encompassing multiple versions of each smart contract in order to cover different ways of satisfying or violating its associated properties.
%
%
To foster the reproducibility of the results, we make available a toolchain that  automatises the construction of the verification tasks, their processing with SolCMC and Certora, and the summarisation of the results.
Based on these artifacts, we present a preliminary evaluation of 
SolCMC and Certora, comparing their completeness, soundness, and expressiveness. 
Finally, we introduce a scoring scheme for Solidity verification tools, which is inspired by schemes used in software verification competitions~\cite{Beyer23tacas}, but taking into account the peculiarities of the smart contracts context.  





%% file: related.tex
\section{Background and related work}
\label{sec:related}


Over the years, several dozens of tools have been developed to analyse Ethereum  contracts (see \eg~\cite{Kushwaha22access,Ivanov23csur,GarfattaKGG21acsw} for systematic surveys).
The vast majority of these tools focus on specific types of contract vulnerabilities, such as reentrancy, integer overflow and underflow, mishandled exceptions, transaction ordering dependence, \etc~\cite{swc}. 
%
More recent tools give users more control on the properties to be verified, in principle enabling the verification of contract implementations against an ideal, abstract description of their behaviour. 


A prominent tool in this category is SolCMC~\cite{Solcmc}, a symbolic model checker integrated in the Solidity compiler since 2019. 
Specifying properties in SolCMC requires  developers to instrument the contract code with \solcode{assert} statements, which are treated as verification targets.
Failure of an \solcode{assert} means that the desired property is not satisfied by the contract.
For example, consider a method \solcode{deposit} that receives ETH from any user, recording the sent amount in a \solcode{balances} mapping.
The property ``after a successful \solcode{deposit}, the balance entry of 
\lstinline{msg.sender} is increased by \solcode{msg.value}'' can be encoded as the function in~\Cref{lst:solcmc:deposit-user-balance}:

\begin{listing}[ht]
\caption{SolCMC encoding of a safety property.}
\label{lst:solcmc:deposit-user-balance}
\begin{tcolorbox}[myoneside]
\centering
\begin{lstlisting}[language=solidity]
function deposit_user_balance() public payable {
  uint old_user_balance = balances[msg.sender];
  deposit();
  uint new_user_balance = balances[msg.sender];
  assert(new_user_balance == old_user_balance + msg.value);
}

\end{lstlisting}
\end{tcolorbox}
\end{listing}

This defines a contract invariant that must be true for any reachable contract state: the balance of a user after a successful call is equal to the previous balance plus the deposit.
SolCMC transforms the instrumented contract into a set of Constrained Horn Clauses (CHC) \cite{Bj&15,DeA&22}
 which is fed to a CHC satisfiability solver (Spacer~\cite{Spacer}, integrated in Z3~\cite{DeB08}, or Eldarica~\cite{Eldarica}) to check if any assert can fail.
If so, it produces a trace witnessing the violation.


Certora~\cite{certora,certora-wip} is another leading formal verification tool for Solidity. 
Unlike SolCMC, it decouples the specification of the properties from the contract code.
Properties, written in the Certora Verification Language (CVL), roughly can take the form of \cvlcode{assert} statements (``for all contract runs, the condition holds'') or \cvlcode{satisfy} statements (``there exists a run where the condition holds'').
For instance, the CVL specification of the \solcode{deposit} property seen before 
is shown in~\Cref{lst:certora:deposit-user-balance}.
Certora compiles the Solidity contract and its associated properties into a logical formula,
and sends it to an SMT solver.
Another key difference between Certora and SolCMC is that in SolCMC verification is done locally (since it is part of the Solidity compiler stack), while in Certora it is executed remotely on a cloud service.



\begin{listing}[ht]
\caption{Certora encoding of a safety property.}
\label{lst:certora:deposit-user-balance}
\begin{tcolorbox}[myoneside]
\centering
\begin{lstlisting}[language=solidity]
rule deposit_user_balance {
  env e; // an arbitrary transaction and context
  address sender = e.msg.sender;
  mathint old_user_balance = getBalanceEntry(sender);
  deposit(e); // calls deposit with context e 
  mathint new_user_balance = getBalanceEntry(sender);
  mathint deposit_amount = to_mathint(e.msg.value);
  assert new_user_balance == old_user_balance + deposit_amount;
}
\end{lstlisting}
\end{tcolorbox}
\end{listing}


Besides SolCMC and Certora, other tools for verifying user-defined properties of Solidity contracts have been proposed 
(see~\cite{Solcmc} for a comparison).  
VerX~\cite{Permenev20sp} models properties in a variant of past linear temporal logic.
This allows to verify safety properties of contracts, while liveness properties are not expressible.
%
%
SmartACE~\cite{Wesley22vmcai} verifies properties written in Scribble~\cite{scribble}: contracts are annotated with Scribble annotations (\ie, contract invariants and method postconditions). 
Scribble transforms the annotated contract into a contract with \solcode{assert}s, which are used as verification targets. 
SmartACE uses local bundle abstractions to reduce the state explosion caused by having to deal with many users integrating with the contract, factorizing users into a representative few. It models each contract in LLVM-IR and integrates existing analysers such as SeaHorn and Klee to facilitate verification. 
Notably, SmartAce has been applied to verify some contracts from the OpenZeppelin library~\cite{smartace-openzeppelin}.


\paragraph*{Comparing verification tools}

A primary source of comparison among different verification tools is given by the research papers where these tools were introduced
\cite{Feist19slither,Tsankov18ccs,Ethor}.
%
%
A problem here is that each comparison is based on an ad-hoc dataset of contracts and properties, which makes it difficult to compare the effectiveness of different tools.
The work~\cite{Angelo23wtsc} provides a unifying view of these datasets, by
collecting their verification tasks and judgements, and mapping them to a uniform scheme based on the Smart Contract Weakness Classification~\cite{swc}. 
A main difference between this dataset and ours is in the nature of the properties in the verification tasks: the ones in~\cite{Angelo23wtsc} are specific vulnerabilities (\eg, reentrancy, overflows, \etc),
while ours are ideal properties of the analysed contract 
(\eg, ``after calling \solcode{foo}, the sender receives 1 ETH'').

A few works compare different analysis tools without introducing their own. 
The work~\cite{Chaliasos24icse} evaluates five tools based on their ability to identify vulnerabilities that have been actually exploited by attacks in the wild.
Perhaps surprisingly, the conclusion is that tools that detect specific vulnerability patterns are ineffective against real attacks, being able to counter only $\sim$12\% to the economic damage in the considered dataset, while offering no protection against the remaining part of the damage, which exceeds 2 billion dollars. This is a strong motivation for research on analysis techniques and tools that can also detect logic-related bugs, which are the focus of our benchmark.
The work~\cite{Dias21prdc} proposes a vulnerability classification scheme that extends~\cite{swc}, and evaluates the effectiveness of three bug detection tools.
%
We note that both works~\cite{Dias21prdc,Chaliasos24icse} focus on tools that detect specific vulnerabilities: at the best of our knowledge, ours is the first comparison between general verification tools for Solidity. Another main difference between our work and \cite{Dias21prdc} is that the comparison in~\cite{Dias21prdc} is based on a quantitative evaluation of the tool outcomes (in the form of a confusion matrix), while we also devise a qualitative comparison that explains the reason behind these results, and in particular the causes of unsoundness (false positives) and incompleteness (false negatives). 

%% file: benchmark.tex
\section{Our benchmark}
\label{sec:benchmark}


The benchmark is logically organized in the following components:
\begin{itemize}

\item a collection of informal specifications of use cases for smart contracts, each accompanied by a set of desirable properties. 
We deliberately choose \emph{not} to use a formal language to write the smart contract specifications and the associated properties, 
since we want to be free to express properties that go beyond those expressible by current verification tools.
 
\item Solidity implementations of the use cases and specifications of their properties in the languages supported by SolCMC and Certora.
For each use case we provide multiple Solidity implementations, either respecting
the given properties or violating them (in obvious or subtle ways).
A \emph{verification task} comprises the implementation of a use case and that of a related property.

\item a \emph{ground truth} that assesses,
for each verification task, whether the implementation satisfies the associated property or not.
\end{itemize}

Our toolchain processes these data to construct the verification task and runs SolCMC and Certora on each of them.


The way we construct the verification tasks is tool-specific: 
\begin{itemize}

\item for SolCMC, each property is encoded in Solidity within the associated smart contract. Although, in general, these asserts can be scattered throughout the contract code, in our benchmark we keep the definitions of the properties separated from the contracts, in order to automatize the verification of multiple properties  on multiple version of the contract.

Accordingly, we provide two ways to write a property: 
    \begin{itemize}
    \item as a function that is added to the contract. This function may assert invariants on the contract state, and may call other contract methods as a property wrapper.  
    \item as a set of fragments of ghost code that are injected in the contract methods.
    \end{itemize}
    In this way, whoever extends the benchmark can write these properties without   
    affecting the behaviour of the original contract, so that the instrumented contract
    satisfies the considered properties if and only if the original one does.
    In practice, this can be achieved by preventing ghost code from writing the state of the original contract and from changing its control flow except to signal the violation of the desired property. 
    Future versions of the toolchain will give warnings when detecting potential discrepancies. 
    
\item for Certora, we write properties in the Certora Verification Language (CVL)~\cite{certora-cvl}.
The syntax of CVL extends 
Solidity with a set of meta-programming primitives that allow to express complex contract properties.  
We encode a contract property as a CVL \emph{invariant} when the property involves facts about the state of the smart contract that should be true in any execution,
while we encode it as a \emph{rule} when the property concerns the expected behavior of calling one or more contract methods. 
In general, a rule is a sequence of commands that describe an execution trace of the contract, together with preconditions (\lstinline{require}) and postconditions.
There are two kinds of postconditions:
\cvlcode{assert}, which \emph{must} hold for any trace, and
\cvlcode{satisfy}, which \emph{can} be satisfied 
(\ie, it is possible to find a trace that makes them true).
Unlike SolCMC, in Certora ghost code can be encoded directly within the properties, without altering the contract being verified.

\end{itemize}

\paragraph*{Scoring the results of the tools}

After constructing the verification tasks, the toolchain runs SolCMC (locally) and Certora (remotely) on each of them.
The execution outcome on a verification task is summarized
and scored according to the schema reported in~\Cref{tab:scoring}.
The overall design goal is that a tool that does nothing will have a null score, 
a tool that provides correct answers when verifying or detecting violation of the properties will have a positive score, and a tool that tricks the user into believing false results 
(\eg, claiming that a property holds when it is not the case \textit{false positive}, or, viceversa, gives a lot of \textit{false negative} answers)  will have a negative score.
We assign a null score in three cases: 
when the property is not expressible in the tool, 
when the tool fails to provide an output
(\eg, because of aborts, timeouts, or memory exhaustion), and
when the tool 
does not provide a definite answer about
the validity of a property. 
The ratio for assigning the same score here is that it would be easy to make the property expressible by a tool that always diverges \robnote{migliorare ora che FN=0}. 

Our viewpoint is that tools are aimed at certifying that desirable properties  
are satisfied by a given Solidity implementation.
Therefore, our scoring schema privileges soundness over completeness, as a false positive may create much bigger problems to users, as they will be convinced that their contract satisfies a property that in practice does not hold, while a false negative will only make the user doubt of the correctness of the contract.

The basis of our scoring schema is standard: 
we have two cases (P/N) depending on whether the property in the verification task
is satisfied or not,
and two cases (T/F) depending on whether the tool answers correctly to the task or not.
We slightly deviate from this standard classification, in that we additionally classify 
the outputs of a tool as \emph{strong} claims 
(\eg, ``the property holds'', ``the property is violated'') 
or \emph{weak} claims
(\eg, ``the property \emph{might} hold'', ``the property \emph{might} be violated'').
More specifically, we use the following criteria to distinguish between weak and strong claims of the tools at hand:
\begin{itemize}

\item in SolCMC, when verification terminates, the output has one of the following forms:
\begin{itemize}
\item ``Assertion violation check is safe!''. We consider this as a strong claim that the tool has verified the property, hence we classify the output as a P!
\item ``Assertion violation happens here''. In this case, the tool outputs the line of code where the asserted property is violated, and shows a sequence of method calls that lead to the violation. Hence, we consider this as a strong claim, and classify it as a N!
\item ``Assertion violation might happen here''. Here the tool has not been able to verify neither the violation, nor the correctness of the property, and (to stay on the safe side) it states that a violation is possible. We classify this output as a N.
\end{itemize}

\item in Certora the classification depends both on the tool output and how the property is modelled in CVL. 
For an \cvlcode{assert}, the output is classified as P! when Certora returns ok, and N when it rejects the property. 
We do not put the ``!'' in the negative case as Certora may reject the property because of an unreachable counterexample, while we put the ``!'' in the positive case because it has been able to prove that no state (reachable or not) leads to a violation.
The case \cvlcode{satisfy} is dealt with dually: we classify as P when Certora returns ok, and N! when it rejects the property.%
\footnote{It is possible to limit the set of considered starting states by refining the implementation with a set of invariants. We do this in a best effort manner.}
Since our benchmark only admits rules that do not use both \cvlcode{assert} and \cvlcode{satisfy}, this criterion is always applicable. 

\end{itemize}



The scoring schema is displayed in~\Cref{tab:scoring}.
Coherently with our design choices, we assign FP! the lowest score,
because the tool is falsely claiming that a desirable property is true.
Strong false negatives FN! (\ie, false alarms) 
are considered ``half as dangerous'' than false positives.  
False weak accepts (FP) have only a mildly negative score, 
since we treat them as mere conjectures:
in fact, when conjecturing a result, the tool is just conveying the fact that it is not convinced of the truth of the opposite result.
The asymmetry between FP! and FN! is mimicked on weak judgements by assigning
false weak rejects (FN) a null score.

\begin{table}
\centering
\small
\caption{Scoring schema for Solidity verification tools.}
\label{tab:scoring}
\begin{tabular}{lcp{9cm}}
\textbf{Result} & \textbf{Points} & \textbf{Description}  
\\
\hline
ND  & 0   & Property not expressible in the tool \\
UNK & 0   & Timeout / Memory exhaustion \\
TN! & 2   & Property violated, tool claims violation \\
TN  & 1   & Property violated, tool conjectures violation \\
FN! & -8  & Property holds, tool claims violation  \\
FN  & 0   & Property holds, tool conjectures violation \\
TP! & 2   & Property holds, tool claims correctness \\  
TP  & 1   & Property holds, tool conjectures correctness \\
FP! & -16 & Property violated, tool claims correctness \\
FP  & -1  & Property violated, tool conjectures correctness \\
\hline
\end{tabular}
\end{table}

%% file: solcmc-certora.tex
\section{Evaluation: SolCMC \emph{vs.} Certora}
\label{sec:solcmc}
\label{sec:certora}

We discuss in this section what we have learnt by using SolCMC and Certora in the design and application of our benchmark.
Here we focus on completeness, soundness and expressiveness of the two tools: 
their scoring as per~\Cref{tab:scoring} is presented at the end of the section.
As a disclaimer, we note that our evaluation is based on the current versions
of the tools%
\footnote{Versions: solc v0.8.24, Eldarica v2.0.9, Z3 v4.12.2, certora-cli v6.3.1}. Since they are moving targets, with multiple updates released during the writing of this paper, it is likely that some of the weaknesses discussed below may be fixed in future releases. 


\paragraph*{Completeness}

SolCMC and Certora share some sources of incompleteness, 
\ie properties that are true but that the provers do not manage to prove.
This is the case \eg for contracts containing external calls, \ie calls from the analysed contract to another account~\cite{solcmc-untrusted-calls}
(\eg, the \href{https://github.com/fsainas/contracts-verification-benchmark/tree/main/contracts/deposit_erc20}{Deposit/ERC20} use case). 
By default, called contracts are considered untrusted by SolCMC and Certora, 
and accordingly these tools over-approximate their behaviour 
(even when their code is known).
This basically makes the provers fail to verify any property that depends on the behaviour of the called contract, so leading to false negatives.
For example, the \solcode{assert} in~\Cref{lst:solcmc:untrusted-external-calls} always passes, but SolCMC detects a possible violation
(the same happens with Certora).
SolCMC has an option to change the default behaviour by considering external calls to be trusted, but a known drawback of this option is a substantial computational overhead.
For instance, checking the invalidity of the assertion \solcode{c.n()==0} in the contract \solcode{C2} of~\Cref{lst:solcmc:untrusted-external-calls} takes 
$\sim 8$ minutes 
using the Z3 solver in our experimental setting, despite the contract being just a few lines of code (by contrast, it takes a few seconds with the untrusted option).
In general, even with the default option about untrusted calls,
non-termination is not uncommon (see~\Cref{tab:scoring:solcmc-certora}).
This is a general problem related to 
Z3, which can be easily misled to divergence
even with apparently harmless sets of constraints.
Even in real-world use cases not specifically crafted to make verification
burdensome, computation times sometimes occur to explode unexpectedly.  

\begin{listing}[ht]
\caption{SolCMC: untrusted external calls.}
\label{lst:solcmc:untrusted-external-calls}
\centering
\begin{tcolorbox}[mytwosides]
\centering
\begin{lstlisting}[language=solidity]
contract C1 {
  uint n; 
  constructor() { n=0; }

  function set() external { n=1; }
  
  // n could be either 0 or 1
}
\end{lstlisting}
\tcblower
\begin{lstlisting}[language=solidity]
contract C2 {
  C1 public c;
  constructor() { c=new C1(); }
    
  function inv() public view {
    assert(c.n()<=1); 
  }
}
\end{lstlisting}
\end{tcolorbox}
\end{listing}

Many desirable properties of accounts, like \eg that the balance is updated according to certain rules, are often broken on \emph{contract} accounts. 
For instance, consider a contract with a method that allows the sender to withdraw funds, and the property ``after a successful call to \solcode{withdraw}, the balance of \solcode{msg.sender} has increased''. 
This property may be violated when the sender is a contract account, which fallbacks on the \solcode{withdraw} by giving away all its balance. 
While properties of this kind do not hold, in general, for contract accounts, they are expected to hold for externally-owned accounts (EOAs), which do not have code
(\eg, \code{withdraw-sender-rcv-EOA} in the \href{https://github.com/fsainas/contracts-verification-benchmark/tree/main/contracts/bank}{bank} use case).
In general, properties of EOAs are not even expressible, since it is not possible to discriminate EOAs from contract accounts
(see the discussion below about expressiveness).
A property about EOAs is expressible only if the account under scrutiny is the \solcode{msg.sender}.
In this case, it is possible to tell that the account is an EOA by comparing it to \solcode{tx.origin} (the transaction originator):
namely, the two addresses are equal iff \solcode{msg.sender} is an EOA.
The property above can then be refined as
``if \solcode{msg.sender} is an EOA, then after a call to \solcode{withdraw}, the balance of \solcode{msg.sender} has increased''.
Both SolCMC and Certora fail however to verify that the amended property is satisfied.
No alternative encodings of EOAs seem to exist that allow the provers to successfully verify non-trivial properties about them.


Further cases of incompleteness include map invariants (\eg, the sum of a map is preserved), which are unlikely to be proved by both tools, and the over-approximation of the possible environments. 
\Eg, Certora includes the contract in the approximation for \solcode{msg.sender}, even if the contract has no calls 
(\eg, \code{deposit-contract-balance} in the \href{https://github.com/fsainas/contracts-verification-benchmark/tree/main/contracts/bank}{bank}).



\paragraph*{Soundness}

\begin{listing}[t]
\caption{Unsoundness in SolCMC: \emph{selfdestruct}.}
\label{lst:solcmc:call-wrapper}
\centering
\begin{tcolorbox}[mytwosides]
\centering
\begin{lstlisting}[language=solidity]
contract CallWrapper is ReentrancyGuard {
  function callwrap(address called) public nonReentrant {
    called.call("");
  }
  ...
}
\end{lstlisting}
\tcblower
\begin{lstlisting}[language=solidity]
// SolCMC invariant
function inv(address a) public {
  uint b = address(this).balance;
  callwrap(a);
  
  // contract balance is preserved 
  assert(b==address(this).balance);
}
\end{lstlisting}
\end{tcolorbox}
\end{listing}

Analysing the results of our benchmark, we have spotted a few sources of unsoundness (\ie, the property does not hold but the tool falsely claims it is true) for both SolCMC and Certora.
In SolCMC, false positives may happen when reasoning about the contract balance in case of external calls and reentrancy guards, as shown in~\Cref{lst:solcmc:call-wrapper} 
(see the \code{bal} property in the \href{https://github.com/fsainas/contracts-verification-benchmark/tree/main/contracts/call-wrapper}{CallWrapper} use case). 
The contract \solcode{CallWrapper} has a single method, which performs a low-level call to an arbitrary address; the \solcode{nonReentrant} modifier by OpenZeppelin ensures that this call is non-reentrant.
Now, consider the property: ``the contract balance is preserved by \solcode{callwrap}''.
Apparently, it might seem to hold, because the contract has no \solcode{payable} nor \solcode{receive} methods.
However, there are other asynchronous events that can make the contract balance increase: \eg, it can receive ETH from a \emph{coinbase} transaction 
(\ie, the first transaction in a block, which collects the block reward),
or from a \emph{selfdestruct}
(\ie, an action performed by a contract to destroy itself and transfer the remaining ETH to another account) \cite{solcmc-contract-balance}.
Therefore, the property does \emph{not} hold, since the address called by \solcode{callwrap} can be a contract that triggers a \emph{selfdestruct}.
SolCMC here produces a false positive, claiming that the invariant \solcode{inv} is always satisfied.
We conjecture that this output derives from an under-approximation of SolCMC, which believes that the absence of reentrant methods implies that the call cannot affect the contract state, including the balance. 
Note instead that, when removing the \solcode{nonReentrant} modifier, SolCMC correctly detects that the invariant may be violated.
Certora instead correctly classifies the property as false, but it produces a false positive on an extension of \solcode{CallWrapper} with a variable \solcode{s} that can be updated by the method \solcode{set}, and the property ``\solcode{s} is preserved by \solcode{callwrap}'' 
(see~\Cref{lst:certora:call-wrapper} and the \code{stor} property in the \href{https://github.com/fsainas/contracts-verification-benchmark/tree/main/contracts/call-wrapper}{CallWrapper} use case).
This property is false, since the account called by \solcode{callwrap} can perform a reentrant call to \solcode{set}.
Certora fails to understand that a call to an address may lead to additional code execution, including a further call to one of the methods of the contract. 
Hence, Certora claims that the property holds, hereby being unsound~\cite{certora-report-callwrapper}.

\begin{listing}[ht]
\caption{Unsoundness in Certora: untracked reentrant calls.}
\label{lst:certora:call-wrapper}
\centering
\begin{tcolorbox}[mytwosides]
\centering
\begin{lstlisting}[language=solidity]
contract CallWrapper {
  uint s;
  ...
  function set(uint snew) public {
    s = snew;
  }
}
\end{lstlisting}
\tcblower
\begin{lstlisting}[language=cvl]
rule P(address a) {
  env e;
  uint s0 = currentContract.s;
  callwrap(e, a);
  uint s1 = currentContract.s;
  assert s1 == s0;
}
\end{lstlisting}
\end{tcolorbox}
\end{listing}

Certora has some further documented under-approximations that may lead to unsoundness~\cite{certora-invariants,certora-approx}. 
When dealing with invariants, Certora checks that the invariant is preserved after the execution of each contract method, neglecting the effect of \emph{selfdestruct} or \emph{coinbase} transactions. 
These transactions may increase the ETH balance of the analysed contract, which has no way of preventing this unexpected incoming ETH: therefore, if the invariant depends on the contract balance, it may be broken at any time.   
As an example, consider the contract \lstinline{DoNothing} in~\Cref{lst:certora:donothing} \robnote{bank/user-balance-inc-onlyif-deposit}, which just saves its initial balance in a variable \lstinline{bal0}, and does nothing afterwards. 
The associated CVL property is an invariant checking that the contract balance is always equal to the stored initial balance. 
Certora claims that the invariant holds, since it holds at creation and it is preserved after the execution of each method (\lstinline{balanceOf}). 
This is unsound, since \eg a \emph{coinbase} transaction can send ETH to the contract, making the actual balance exceed \lstinline{bal0}.

\begin{listing}[ht]
\caption{Unsoundness in Certora: untracked \emph{selfdestruct} and \emph{coinbase} transactions.}
\label{lst:certora:donothing}
\centering
\begin{tcolorbox}[mytwosides]
\centering
\begin{lstlisting}[language=solidity]
contract DoNothing {
  uint bal0;
  constructor() {
    bal0 = address(this).balance;
  }
  function balanceOf(address a) public view returns (uint) {
    return a.balance;
  }
}
\end{lstlisting}
\tcblower
\begin{lstlisting}[language=cvl]
// Certora invariant specification

invariant inv()
  balanceOf(currentContract) 
  == 
  currentContract.bal0;
\end{lstlisting}
\end{tcolorbox}
\end{listing}

Besides the artificial example in~\Cref{lst:certora:call-wrapper}, false positives may occur in real-world use cases when reasoning about the state after a low-level call.
For instance, in our benchmark this is the case for a simple
\href{https://github.com/fsainas/contracts-verification-benchmark/tree/main/contracts/bank}{bank} contract allowing users to deposit and withdraw ETH, and the property requiring that after a successful \solcode{withdraw}, the balance entry of the sender is decreased of the right amount.
Apart from these glitches, both SolCMC Certora perform comparably well regarding false positives on our benchmark (see~\Cref{tab:scoring:solcmc-certora}).


\paragraph*{Expressiveness limitations}

One of the main categories of properties that cannot be encoded in SolCMC are liveness properties 
(\eg, \code{wd-fin-before} in the \href{https://github.com/fsainas/contracts-verification-benchmark/tree/main/contracts/vault}{vault} use case). 
For a minimal example, consider \Cref{lst:solcmc:liveness} and the property ``\lstinline{foo} never reverts''. 
Intuitively, we can call \lstinline{foo} inside of a \solcode{try-catch} statement, making sure that the method does not revert by checking that the \solcode{catch} is not reachable. However, this encoding is unsound, since the invariant is satisfied by some implementations of \lstinline{foo} that actually revert. Specifically, this is the case of implementations of \lstinline{foo} that never revert when the method is called by the contract itself, like the one in~\Cref{lst:solcmc:liveness}, left. 
The invariant passes, but if foo is called from any account different from \lstinline{Liveness}, then \lstinline{foo} reverts.
This is not the only way to trick SolCMC into verifying a false liveness property, as any assumption made by having the contract call itself can be used (\eg, reentrancy).
In general, when expressing a property in SolCMC, any external call made in the invariant does not faithfully capture the intended property, as it does not model a call made by an arbitrary user, but only by the contract itself. 
The same problem exists when we use a low-level call instead of an external call within a \solcode{try-catch}.
Attempting to encode the liveness property as the success of the invariant itself does not work either: any command in the body of the invariant cannot ensure that a certain line of it is always reached. 

\begin{listing}[ht]
\caption{Completeness in SolCMC: a wrong encoding of a liveness property.}
\label{lst:solcmc:liveness}
\centering
\begin{tcolorbox}[mytwosides]
\centering
\begin{lstlisting}[language=solidity]
contract Liveness {
  function foo() {
    require(msg.sender==
            address(this));
  }
  ...
}
\end{lstlisting}
\tcblower
\begin{lstlisting}[language=solidity]
// wrong encoding of "foo succeeds"

function inv() public {
  try this.foo() {} catch {
    assert(false);
  }
}
\end{lstlisting}
\end{tcolorbox}
\end{listing}

Even restricting to safety, expressing properties that involve two or more method calls in sequence is tricky, and in particular it cannot be done by using invariants, only.
\Eg, the invariants in~\Cref{lst:solcmc:sequence} (right) are a seemingly reasonable (but unsound) encoding of the property ``\lstinline{bar} cannot be called twice in a row''
(see also the property \code{wd-twice} in the \href{https://github.com/fsainas/contracts-verification-benchmark/tree/main/contracts/vault}{vault}).

Consider \eg the  implementation of \lstinline{bar} in the contract \lstinline{Sequence}: here, two consecutive calls to \lstinline{bar} are possible whenever the callers are distinct: hence, the property is violated.   
Notice instead that the invariant is satisfied, because the sender of both calls to \lstinline{bar} is the same, since coincides with the sender of \lstinline{inv}.
A sound encoding is still possible, but at the cost of instrumenting the contract with ghost code, which although supported by our toolchain, is a complex and error-prone operation in general.  
By contrast, Certora can express smoothly this kind of properties as CVL rules, whenever the number of calls in the sequence is fixed.
Properties involving \emph{unbounded} sequences are instead not expressible even in Certora 
(see \eg the property \code{always-wd-all-many} in the \href{https://github.com/fsainas/contracts-verification-benchmark/tree/main/contracts/zerotoken_bank}{tokenless bank} use case).

\begin{listing}[ht]
\caption{Completeness in SolCMC: multiple sequential calls.}
\label{lst:solcmc:sequence}
\centering
\begin{tcolorbox}[mytwosides]
\centering
\begin{lstlisting}[language=solidity]
contract Sequence {
  address last;
  function foo() { ... }
  function bar() {
    require(msg.sender!=last);
    last = msg.sender;
  }
  ...
}
\end{lstlisting}
\tcblower
\begin{lstlisting}[language=solidity]
function inv() public {
  bar(); 
  bar();
  assert(false);
}
\end{lstlisting}
\end{tcolorbox}
\end{listing}


The property specification language supported by Certora is quite powerful, allowing us to express most of the properties in our benchmark.
Still, there are interesting classes of properties that cannot be expressed. 
This is the case \eg for properties of the form ``for all reachable states, some user can do something that eventually produces some desirable effect''.
For example, consider the \lstinline{Deposit} contract in~\Cref{lst:certora:liquid-nd}.
The contract allows anyone to withdraw any fraction of its balance through the method \lstinline{pay}, unless the variable \lstinline{frozen} is true.
Now, \lstinline{frozen} is controlled by the contract owner through the method \lstinline{freeze}: hence, the owner at any point can freeze the contract balance, preventing anyone from withdrawing.
A desirable property contracts, in general, is that the funds stored in the contract cannot be frozen forever, a property often referred to as \emph{liquidity}~\cite{Tsankov18ccs,BMZ22lmcs,Laneve23jlap}.
A tentative formalization of the liquidity property is the CVL rule in~\Cref{lst:certora:liquid-nd}. 
The rule is satisfied if there exists some starting state such that,
for all \lstinline{sender} address and value \lstinline{v},
\lstinline{sender} can fire a transaction \lstinline{pay(v)} that increases their balance by \lstinline{v}.
This however is not a correct way to encode our liquidity property: indeed, Certora says that the property is satisfied, since there \emph{exists} a trace that makes the condition in the \lstinline[language=cvl]{satisfy} statement true (this is the trace where the owner has not set \lstinline{frozen}). 
Note that an alternative formalization where the \lstinline[language=cvl]{satisfy} is replaced by an \lstinline[language=cvl]{assert} does not work too.
In this case, we would require that, for all reachable states, a transaction 
\lstinline{pay(v)} is never reverted, \emph{for all} choices of the amount \lstinline{v}.
Certora would correctly state that the property is false, because there are some values \lstinline{v} that make the transaction fail (\eg, when \lstinline{v} exceeds the contract balance). 
Although in this simple case it would be easy to fix the property by requiring that the transaction is not reverted for all values \lstinline{v} less than the contract balance and when \lstinline{frozen} is false, 
in general for liquidity we would like to know if there \emph{exist} parameters that make the desirable property true, which is not expressible in CVL.


\begin{listing}[ht]
\caption{Expressiveness of Certora: along all paths, eventually.}
\label{lst:certora:liquid-nd}
\centering
\begin{tcolorbox}[mytwosides]
\centering
\begin{lstlisting}[language=solidity]
contract Deposit {
  address owner;
  bool frozen;
  constructor () payable {
    owner = msg.sender;
  }
  function freeze() public {
    require (msg.sender == owner);
    frozen = true;
  }
  function pay(uint v) public {
    require(!frozen);
    (bool succ,) = msg.sender.call{value: v}("");
    require(succ);	
  }
}
\end{lstlisting}
\tcblower
\begin{lstlisting}[language=cvl]
// Certora rule specification

rule P(address sender, uint v) {
  // sender initial balance
  mathint b0 = bal(sender);
  env e;

  require e.msg.sender == sender;

  pay(e, v);

  // sender balance after pay(v)
  mathint b1 = bal(sender);

  // looking for a positive example
  satisfy(b1 == b0 + v);
}
\end{lstlisting}
\end{tcolorbox}
\end{listing}

Another category of properties that are not easily expressible, are those that reason about transfers of funds from the contract. For example, consider the property ``the method \solcode{pay} calls the sender, transferring 10  wei from the contract to it'', which is trivially satisfied by the contract in~\Cref{lst:solcmc:pay-nd}
(see also \code{arbitrate-send} in the \href{https://github.com/fsainas/contracts-verification-benchmark/tree/main/contracts/escrow}{escrow} use case).
It might seem reasonable to express this property as a simple check on the balance of the contract and of the sender, as the invariant in~\Cref{lst:solcmc:pay-nd} (right).
This however would not be correct: in fact, a contract that sends 10 wei to a middle man who forwards the sum to the sender would satisfy the invariant but violate the property.
Indeed, we do not believe that properties of this kind are expressible in SolCMC. 
In CVL instead we can express a property very similar to the above, in which the required called contract is not the sender, but a specific account. This is possible with the use of hooks (this would require using hooks).   
It is unclear whether the exact property above is expressible in Certora.

As mentioned before, neither SolCMC nor Certora can, in general, express properties that are specific to EOAs.
This derives from the fact that EOAs are not always discernible from contract accounts~\cite{openzeppelin-eoa}.
Other classes of properties that seem beyond the expressiveness boundaries of  existing Solidity verification tools are those about game-theoretic interactions between users and adversaries, and fairness and probabilistic properties 
(see, respectively, the properties \code{rkey-no-wd} and \code{okey-rkey-private-wd} in the \href{https://github.com/fsainas/contracts-verification-benchmark/tree/main/contracts/vault}{vault} use case).


\begin{listing}[ht]
\caption{Expressiveness of SolCMC: a wrong attempt of reasoning about received ETH.}
\label{lst:solcmc:pay-nd}
\centering
\begin{tcolorbox}[mytwosides]
\centering
\begin{lstlisting}
contract Deposit {
  // pay 10 wei to the sender
  function pay() public {
    (bool succ,) = msg.sender.call
                   {value: 10}("");
    require(succ);
  }
}
\end{lstlisting}
\tcblower
\begin{lstlisting}
function invariant() public {
  uint s0 = msg.sender.balance;
  uint c0 = address(this).balance;
 
  pay();
  
  uint s1 = msg.sender.balance;
  uint c1 = address(this).balance;
  assert(s1==s0+10 && c1==c0-10);
}
\end{lstlisting}
\end{tcolorbox}
\end{listing}

\paragraph*{Scoring}

We display in~\Cref{tab:scoring:solcmc-certora} the results obtained by running SolCMC and Certora on the verification tasks in our benchmark.
As expected, Certora is able to express more properties than SolCMC; despite that, SolCMC's score is close to Certora's, which is penalized by its seemingly better behaviour with respect to soundness, the causes of which are exactly those discussed before in~\Cref{lst:certora:call-wrapper,lst:certora:donothing}.
Regarding the two CHC solvers used by SolCMC, we note that Z3 has several UNK entries more than Eldarica due to its propensity to timeout. This explains the difference in score between Z3 and Eldarica: the latter performs worse because of its tendency to terminate with the wrong answer (getting -16 for the FP!), whereas Z3 just diverges (getting 0 for the UNK).
Besides that, we do not note significant differences between these two CHC solvers.
We highlight the absence of weak false positives (FP) in our results: SolCMC never claims that a property holds unless it is sure of it, while Certora only does so when the properties are encoded using \cvlcode{satisfy} statements (which have not been used in our current use cases). 

\begin{table}[ht]
\caption{Scoring of SolCMC and Certora (\vertasks verification tasks).}
\label{tab:scoring:solcmc-certora}
\centering
\begin{tabular}{lccccccc}
& \textbf{ND} & \textbf{UNK} & \textbf{TP(!)} & \textbf{TN(!)} & \textbf{FP(!)} & \textbf{FN(!)} & \textbf{Score} 
\\
\hline
\textbf{Certora}           &  60&   0 & 97(97) &  94(0) & 7(7) &  65(0) & 176
\\
\textbf{SolCMC (Z3)}       & 153 & 22 & 86(86) & 49(48) & 2(2) &  11(9) & 165
\\
\textbf{SolCMC (Eldarica)} & 153 &  5 & 88(88) & 54(52) & 7(7) & 16(10) &  90
\\
\hline
\end{tabular}
\end{table}

%% file: conclusions.tex
\section{Conclusions}
\label{sec:conclusions}


We have discussed an ongoing collaborative effort towards the construction of a benchmark for comparing formal verification tools for Solidity. Once completed, the benchmark will serve as a valuable resource for developers, providing guidance in choosing the appropriate tool based on project requirements and contract complexity. 
In the meanwhile, we have given a first qualitative evaluation of SolCMC and Certora by discussing their soundness, completeness and expressiveness limitations based on our experience on developing the benchmark.


Constructing the ground truth was perhaps the most difficult task we encountered while developing the benchmark.
In particular, while it is often easy to tell that an intentionally bugged contract violates a property, ensuring that a property is satisfied is error-prone.
Indeed, Solidity has a few semantical quirks that make manual reasoning about contracts quite burdensome (reentrancy, in particular, may be very tricky).
For this reason, it could make sense to simplify the construction of the ground truth by reducing the number of verification tasks for each use case: for instance, we could provide just a contract version that satisfies all the desirable properties, and just one version witnessing the violation of each property. 
While we are not able to certify the correctness of our ground truth beyond any reasonable doubt, we expect that the open-source nature of our benchmark can foster the collaboration with the community of experts.  
To consolidate our ground truth (at least, for negative properties), we plan to add, in future versions of the benchmark, references to actual contracts in the Ethereum testnet that violate the property.
In this regard, we note that the counterexamples outputted by SolCMC and Certora when they find a violation are rarely informative about its causes.   
Extending the benchmark with more use cases, including more convoluted ways to violate properties, and experimenting it on other verification tools beyond SolCMC and Certora are also viable directions for future work.

